# Algorithmic Complexity of Real Financial Markets.

R. Mansilla[1,2]

[1] Department of Complex Systems, Physical Institute, National University of Mexico, p. o. box 20-364, Del. A. Obregón, 01000, Mexico, D. F.

[2] Faculty of Mathematics and Computer Science University of Havana, Cuba.

## Abstract.

A new approach to the understanding of the complex behavior of financial markets index using tools from thermodynamics and statistical physics is developed. Physical complexity, a magnitude rooted in the Kolmogorov-Chaitin theory is applied to binary sequences built up from real time series of financial markets indexes. The study is based on NASDAQ and Mexican IPC data. Different behaviors of this magnitude are shown when applied to the intervals of series placed before crashes and in intervals when no financial turbulence is observed. The connection between our results and The Efficient Market Hypothesis is discussed.

**PACS numbers:** 05.40.-a, 05.65.+b, 89.90.+n, 02.50.-r

# 1 Introduction.

In the last years financial markets have received a growing attention from general public. Weather storms are discussed with the same emphasis in journals, newspapers and TV news services than the financial ones and their endurance and consequences are analyzed by specialists in both fields.

Physicists have started few years ago to investigate financial data since they are remarkably well-defined complex systems, continuously monitored down to time scale of seconds. Besides, almost every economic transaction is recorded, and an increasing amount of economic data is becoming accessible to the interested researchers. Hence, financial markets are extremely attractive for researchers aiming to understand complex systems.

Several toy models of the markets behavior have been developed, highlighting among them the so-called Minority Game [1]. In that model, a magnitude resembling the real volatility (and called also *volatility*), have been extensively studied [2-9]. As we have proved in [10] there are more sensitive measures of the behavior of the model, which have their origin in statistical physics and thermodynamics. In further papers [11-12] we studied in more details one of these measures called *physical complexity* [13-15]. In [12] we proposed an *ansatz* of the type $C(l) = l^{\alpha}$ for the average value of physical complexity taken over an ensemble of binary sequences. We also proved that the exponent $\alpha$ strongly depends on the parameters of the model.

The aim of this work is to extend the study of this magnitude and his properties to the time series of real financial markets. As we will prove, the behavior of this measure of complexity drastically varies when applied to those intervals of the financial time series placed before the crashes and those where no financial turbulence was observed. We claim that this fact is close related with the Efficient Market Hypothesis (EMH).

The structure of this paper is as follows: In Sec. 2 we discuss the way in which a real time series is encoded in a binary string. We also briefly develop the theoretical tools used in our analysis. In Sec. 3 we expose our results concerning the behavior of the physical complexity and in Sec. 4 we discuss our result in relation with the most accepted paradigm of financial market behavior.

## 2 The binary string associated to the real time series.

One could conceive a financial market as a set of $N$ agents each of them taking a binary decision every time step. This is an extremely crude representation, but capture the essential feature that decision could be coded by binary symbols (buy $=0$, sell $=1$, for example). Although the extreme simplification[1], the above setup allow a "stylized" definition of price.

Let $N_0^t$, $N_1^t$ be the number of agents taking the decision $0,1$ respectively at the time $t$. Obviously, $N = N_0^t + N_1^t$ for every $t$. Then, with the above definition of the binary code the price can be defined as:

$$p^t = f\left(\frac{N_0^t}{N_1^t}\right)$$

where $f$ is an increasing and convex function which also hold that:

a) $f(0) = 0$

b) $\lim_{x \to \infty} f(x) = \infty$

c) $\lim_{x \to \infty} f'(x) = 0$

The above definition perfectly agree with the common believe about how offer and demand work. If $N_0^t$ is small and $N_1^t$ large, then there are few agents willing to buy and a lot of agents willing to sale, hence the price should be low. If on the contrary, $N_0^t$ is large and $N_1^t$ is small, then there are a lot of agents willing to buy and just few agents willing to sale, hence the price should be high. Notice that the winning choice is related with the minority choice. A complete study about the suitability of the properties of the above function $f$ and their consequence will appear in [17].

We exploit the above analogy to construct a binary time series associated to each real time series of financial markets. The procedure is as follows: Let $\{p_t\}_{t \in N}$ be the original real time series. Then we construct a binary time series $\{a_t\}_{t \in N}$ by the rule:

$$a_t = \begin{cases} 1 & p_t \geq p_{t-1} \\ 0 & p_t < p_{t-1} \end{cases}$$

A similar technique was originally introduced in the context of natural languages [18] and more recently in the study of cross-correlation between different market places [19]. To the outgoing binary series we apply the theoretical tools, which we describe below.

Physical complexity (first studied in [13] and [14]) is defined as the number of binary digits that are *explainable* (or meaningful) with respect to the environment in a string $\eta$. In reference to our problem the only physical record one gets is the binary string built up

---

[1] Because among the other reasons, often the agents do not participate and prefer wait

from the original real time series and we consider it as the environment $\varepsilon$. We study the physical complexity of substrings of $\varepsilon$. The comprehension of their complex features has high practical importance. The amount of data agents take into account in order to elaborate their choice is finite and of short range [20]. For every time step $t$ the binary digits $a_{t-l}, a_{t-l+1}, \ldots, a_{t-1}$ represent in some sense the winning choices made by the agents in the corresponding instant of time. Therefore, the binary strings $a_{t-l}a_{t-l+1}\cdots a_{t-1}$ carry some information about the behavior of agents. Hence, the complexity of these finite strings is a measure of how complex information agents face. We study the complexity of statistical ensembles of these substrings for several values of $l$.

We briefly review some measures devoted to analyze the complexity of binary strings. The Kolmogorov - Chaitin complexity [21], [22] is defined as the length of the shortest program $\pi$ producing the sequence $\eta$ when run on universal Turing machine T:

$$K(\eta) = min\left\{|\pi| : \eta = T(\pi)\right\} \qquad (1)$$

where $|\pi|$ represent the length of $\pi$ in bits, $T(\pi)$ the result of running $\pi$ on Turing machine $T$ and $K(\eta)$ the Kolmogorov-Chaitin complexity of sequence $\pi$. In the framework of this theory, a string is said to be *regular* if $K(\eta) < |\eta|$. It means that $\eta$ can be described by a program $\pi$ with length smaller than $\eta$ length.

As we have said, the interpretation of a string should be done in the framework of an environment. Hence, let imagine a Turing machine that takes the string $\varepsilon$ as input. We can define the conditional complexity $K(\eta/\varepsilon)$ [13-15] as the length of the smallest program that computes $\eta$ in a Turing machine having $\varepsilon$ as input:

---

for a more clear position of the market [16].

$$K(\eta / \varepsilon) = min \{|\pi| : \eta = C_T(\pi, \varepsilon)\} \qquad (2)$$

We want to stress that $K(\eta / \varepsilon)$ represents those bits in $\eta$ that are random with respect to $\varepsilon$ [13]. Finally, the physical complexity can be defined as the number of bits that are meaningful in $\eta$ with respect to $\varepsilon$:

$$K(\eta : \varepsilon) = |\eta| - K(\eta / \varepsilon) \qquad (3)$$

Notice that $|\eta|$ also represent (see [13-15]) the unconditional complexity of string $\eta$ i.e., the value of complexity if the input would be $\varepsilon = \emptyset$. Of course, the measure $K(\eta : \varepsilon)$ as defined in Eq. 3 has few practical applications, mainly because it is impossible to know the way in which information about $\varepsilon$ is encoded in $\eta$. However (as shown in [15] and references therein), if a statistical ensemble of strings is available to us, then the determination of complexity becomes an exercise in information theory. It can be proved that the average values $C(|\eta|)$ of the physical complexity $K(\eta : \varepsilon)$ taken over an ensemble $\Sigma$ of strings of length $|\eta|$ can be approximated by:

$$C(|\eta|) = \langle K(\eta : \varepsilon) \rangle_\Sigma \cong |\eta| - K(\Sigma / \varepsilon) \qquad (4)$$

where:

$$K(\Sigma / \varepsilon) = -\sum_{\eta \in \Sigma} p(\eta / \varepsilon) \log_2 p(\eta / \varepsilon) \qquad (5)$$

and the sum is taking over all the strings $\eta$ in the ensemble $\Sigma$. In a population of $N$ strings in environment $\varepsilon$, the quantity $\frac{n(\eta)}{N}$, where $n(s)$ denotes the number of strings equal to $\eta$ in $\Sigma$, approximates $p(\eta / \varepsilon)$ as $N \to \infty$.

Let $\varepsilon = \{a_t\}_{t \in N}$ and $l$ a positive integer, $l \geq 2$. Let $\Sigma_l$ the ensemble of sequences of length $l$ built up by a moving window of length $l$ i.e., if $\eta \in \Sigma_l$ then $\eta = a_i a_{i+1} \cdots a_{i+l-1}$ for some value of $i$. We calculate the values of $C(l)$ using this kind of ensemble $\Sigma_l$. The selection of strings $\varepsilon$ that we do in Sec. 3 is related to periods before crashes and in contrast, period with low uncertainty in the market.

## 3 The study of real time series.

We have used the daily closing value of the NASDAQ composite in the time period from January 3 1995 to April 18 2000 and also the daily closing value of IPC, the leader index of Mexican Stock Exchange (BMV) in the time period from January 2 1991 to March 27 2000. The evolution of both indices can be seen in Fig. 1. We select for both indices several time intervals with the following characteristics:

a) Intervals just before the crashes: the initial point is selected after the onset of the bubble and the last point is that of the all-time high of the index.

b) Intervals where no financial turbulence is observed.

For the NASDAQ we select three periods: from October 13 1995 to May 14 1997, from December 14 1998 to October 28 1999 and from October 28 1999 to February 24 2000. We labeled these periods as N1, N2, N3 respectively. In the period N1 no financial turbulence was observed, N2 correspond to some important season of the Microsoft trial and N3 is the just previous interval to the crash of April 2000 when NASDAQ loss about 25%.

For the IPC we also select three periods: from January 6 1994 to March 6 1995, from January 9 1996 to August 13 1997 and from August 13 1997 to October 25 1999. We labeled these periods as I1, I2, I3. The period N1 was disastrous for Mexican financial market due to the crisis brought about by the presidential transition[2]. In the period N2 no financial uncertainty was observed in economy and period N3 was highly turbulent due to the Asian crisis. We would like to stress the difference between N1 (endogenous origin of the crisis) and N3 (exogenous origin of the crisis).

a)

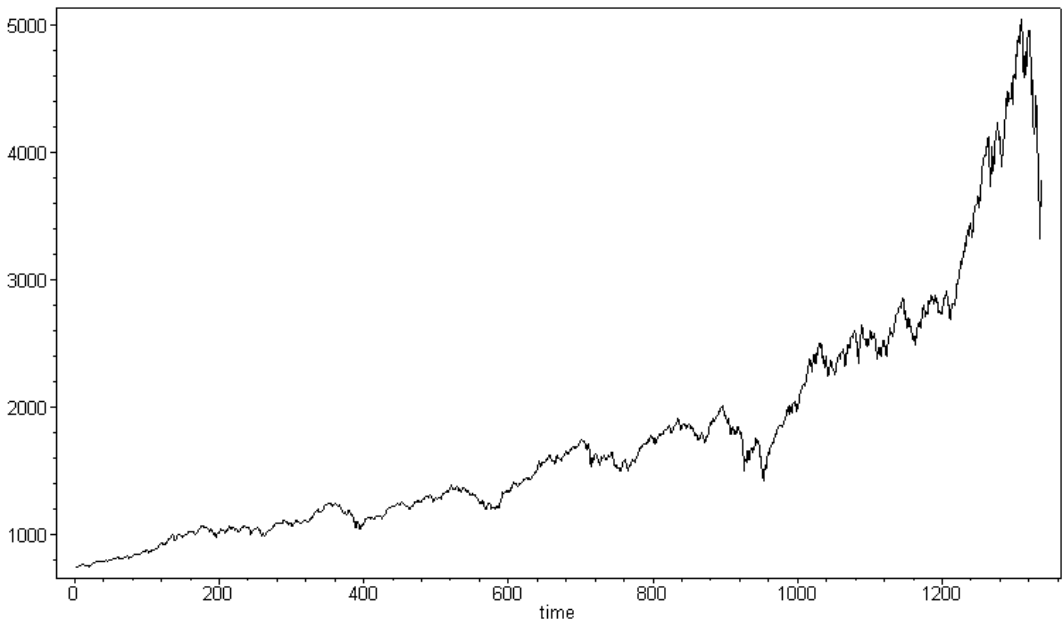

---

[2] This was called "tequilazo".

**b)**

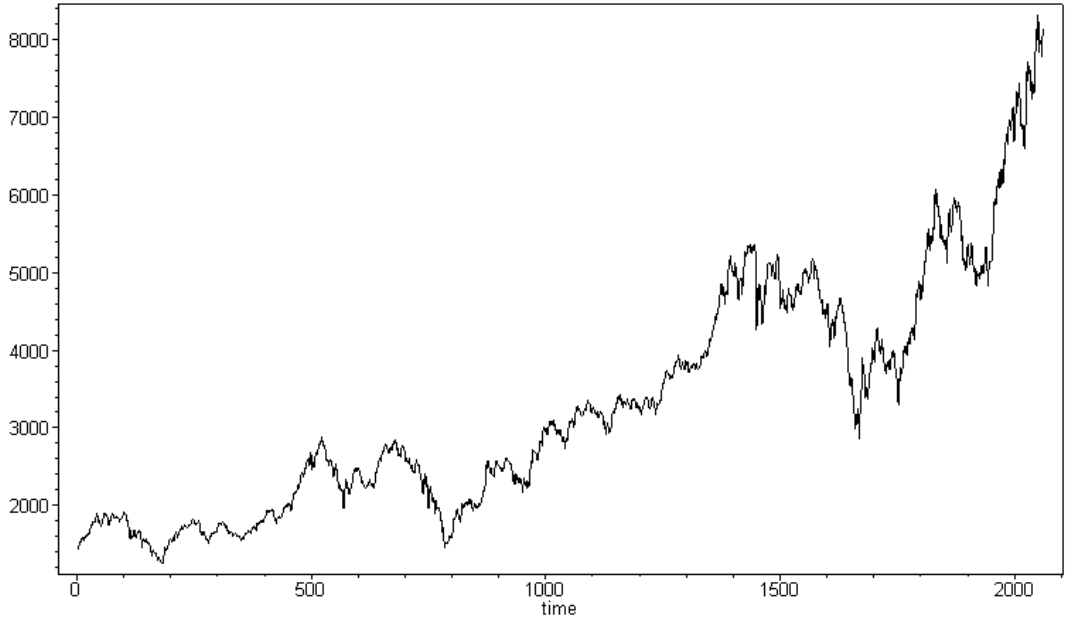

Fig. 1: The evolution of the NASDAQ (a)) index and the IPC (b)) index in the time period from January 1995 to April 2000 and January 1991 to March 2000 respectively. The intervals under study are: N1=(200-360), N2=(1000, 1220), N3=(1220,1300), I1=(500, 790), I2=(180, 500) and I3=(1400, 1750).

**a)**

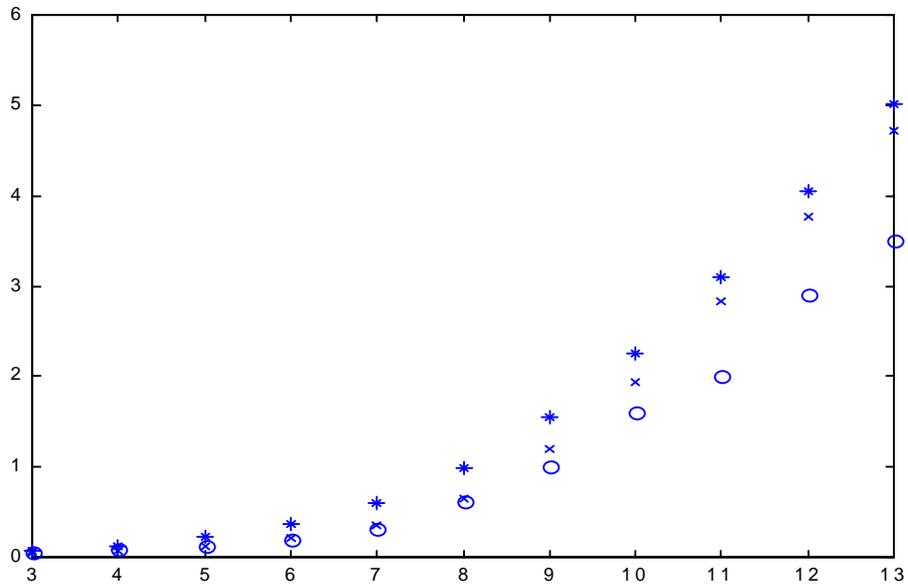

b)

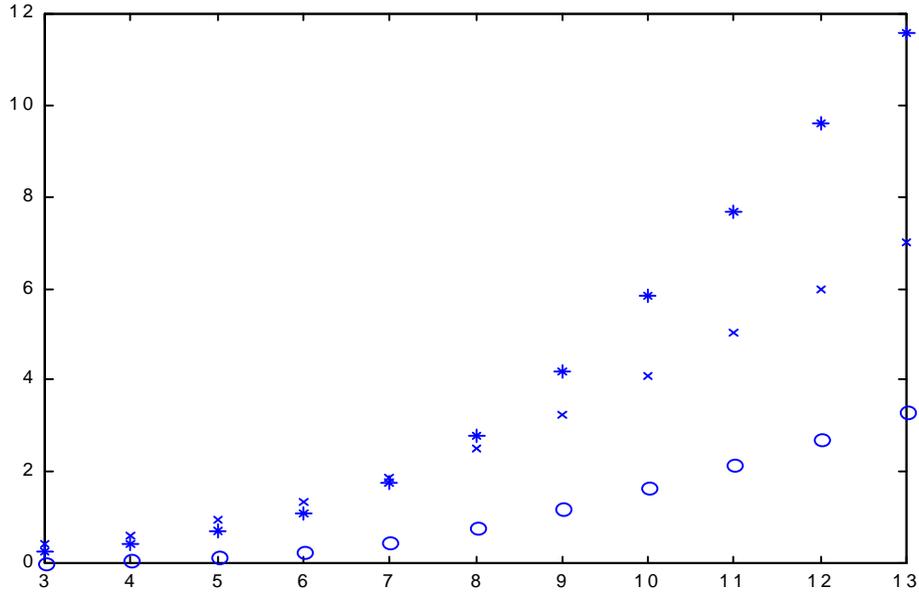

Fig. 2: Values of $C(l)$ vs. $l$ for the intervals of interest (see text). In Fig. 2a we have I1(*), I2=(o) and I3=(x). In fig. 2b we have N1=(o), N2=(x), N3=(*).

We calculate $C(l)$ for the binary sequences associated to the above-mentioned intervals. The results appear in Fig. 2. Notice that the values of the function $C(l)$ corresponding to those intervals where no disturbance is observed (N1 and I2) are lower than those placed just before the crashes (N2, N3, I1 and I3). It means that sequences corresponding to critical periods have more binary digits meaningful with respect to the whole series than those sequences corresponding to periods where nothing happened. If we compare the Fig. 2b with the Fig. 1 of [10] we conclude that N1 behave almost as a random sequence.

In the paper [12] we proposed an *ansatz* of the type $C(l) = \delta\, l^{\alpha}$. The corresponding values of $\delta$ and $\alpha$ for the sequences N1, …, I3 appear in the following Table:

## Table I

## Values de $\delta$ and $\alpha$ for the selected sequences.

| Sequence | $\alpha$ | $\delta$ |
|---|---|---|
| N1 | 2.1059 | 0.0393 |
| N2 | 3.2458 | 0.0014 |
| N3 | 3.6715 | 0.0063 |
| I1 | 3.7748 | 0.0003 |
| I2 | 3.2213 | 0.0013 |
| I3 | 3.6901 | 0.0004 |

The larger exponents correspond to sequences with high financial turbulence. Notice that shorter exponents are related with the more random sequences.

## 4 Conclusions.

The results of the last Section suggest that the intervals where the markets work fine produce binary sequences with features close to the random ones, meanwhile intervals with high uncertainty produce binary sequences, which carry more information.

The above is in agreement with the EMH [23], which stated that the markets are highly efficient in the determination of the most rational price of the traded assets. We conclude that a measure of how close is a markets to the ideal situation described by the EMH is the exponent $\alpha$. The lower exponent, the more random sequence.

More surprisingly is the fact that intervals with high financial turbulence have high informational contents. It open the challenges of understand what kind of information the sequences bear and how it would be used to predict the crashes.

## References.

[1] For a collection of papers and preprints on Minority Game see the web site:

http://www.unifr.ch/econophysics.

[2] D. Challet, Y.C. Zhang, Phys. A **246**, 407 (1997).

[3] Y.-C. Zhang, Europhys. News **29**, 51 (1998).

[4] R. Savit *et al.*, Phys. Rev. Let. **82**, 2203 (1999).

[5] R. Savit *et al*., http://xxx.lanl.gov/abs/adap-org/9712006.

[6] N.F. Johnson *et al*., Phys. A **256**, 230 (1998).

[7] A. Cavagna, http://xxx.lanl.gov/abs/cond-mat/9812215.

[8] N.F. Johnson *et al*., http://xxx.lanl.gov/abs/cond-mat/9903164.

[9] D. Challet, Y.-C. Zhang, Phys. A **256**, 514 (1998).

[10] R. Mansilla, Phys. Rev. E, **61**, 4 (2000).

[11] R. Mansilla, Comp. Sist., **11**, 387 (2000).

[12] R. Mansilla, Phys. A (to be published). See also:

http://xxx.lanl.gov/abs/cond-mat/0002331.

[13] W.H.Zurek, Nature **341**,119 (1984).

[14] W.H.Zurek, Phys. Rev. A **40**, 4731 (1989).

[15] C.Adami,N.J.Cerf, http://xxx.lanl.gov/abs/adap-org/9605002.

[16] J. D. Farmer, J. Ec. Beh. Org. (2000). See also:

http://xxx.lanl.gov/abs/adap-org/9812005.

[17] R. Mansilla, (unpublished).

[18] G. K. Zipf, in Human Behavior and the Principle of Least Effort (Addison-Wesley, Cambridge, MA, 1949).

[19] N. Vandewalle, Ph. Boverox, F. Brisbois, Eur. Phys. J. B (to be published). See also:

http://xxx.lanl.gov/abs/cond-mat/0001293.

[20] B. G. Malkiel, A Random Walk Down Wall Street, $7^{th}$ edn. (W. W. Norton & Co., New York, 1999).

[21] A.N.Kolmogorov, Rus. Math. Sur. **38**, 29-40 (1983).

[22] G.J.Chaitin, J. ACM **13**, 547 (1966).

[23] P. A. Samuelson, Ind. Man. Rev. **6**, 41 (1965).